\documentclass[11pt,a4paper]{article}
\usepackage{latexsym,makeidx,syntonly,amsfonts,amssymb}
\newcommand {\cB}{{\cal B}}
\newcommand {\cC}{{\cal C}}
\newcommand {\cD}{{\cal D}}

\newcommand {\cF}{{\cal F}}
\newcommand {\cG}{{\cal G}}

\newcommand {\cL}{{\cal L}}
\newcommand {\cM}{{\cal M}}

\newcommand {\cO}{{\cal O}}

\newcommand {\cS}{{\cal S}}

\newcommand {\cU}{{\cal U}}

\newcommand {\cW}{{\cal W}}

\newcommand {\cY}{{\cal Y}}
\newcommand {\bcY}{{\ovl{\cal Y}}}
\newcommand {\cZ}{{\cal Z}}

\newcommand{\bl}{{\ovl{ l}}}

\newcommand{\bZ}{{\ovl{ Z}}}

\def\a{\alpha}
\def\b{\beta}

\def\L{\Lambda}

\newcommand{\ovl}[1]{\overline{#1}}

\newcommand{\bz}{\ovl{z}}
\newcommand{\by}{\ovl{y}}                        %new
\newcommand{\zbzk}{(z^k,\ovl{z^k})}
\newcommand{\zbz}{(z,\ovl{z})}
\newcommand{\zbzp}{(z',\ovl{z}')}

\newcommand{\zt}{(\zbz,t)}
\newcommand{\mub}{\ovl{\mu}}
\newcommand{\lam}{\lambda}
\newcommand{\lamb}{\ovl{\lambda}}
\newcommand{\ba}{\ovl{\alpha}}
\newcommand{\bb}{\ovl{\beta}}
\newcommand{\bc}{\ovl{c}}
\newcommand{\bcZ}{\ovl{\cZ}}

\newcommand{\prt}{\partial}
\newcommand{\bprt}{\ovl{\partial}}
\newcommand{\jb}{{\ovl{\jmath}}}

\newcommand{\bby}{\mbox{\boldmath $y$}}
\newcommand{\obby}{\mbox{\boldmath $\ovl{y}$}}

\newcommand{\zY}{(z,\cY)}

\newcommand{\YBY}{(\cY,\ovl{\cY})}
\newcommand{\ZBZz}{(\cZ^\a\zbz,\ovl{\cZ^\a}\zbz)}

\newcommand{\bj}{{\ovl{\jmath}}}
\newcommand{\bi}{{\ovl{\imath}}}
\newcommand{\br}{{\ovl{r}}}
\newcommand{\Sb}{{\ovl{S}}}
\newcommand{\Qb}{{\ovl{Q}}}
\newcommand{\bk}{{\ovl{k}}}
\newcommand{\blambda}{\ovl{\lambda}}
\newcommand{\pt}{\! \cdot \!}

\newcommand{\lra}{\longrightarrow}                           %new

\newcommand{\sect}[1]{\setcounter{equation}{0}\section{\boldmath#1}}

\newcommand{\be}{\begin{equation}}
\newcommand{\ee}{\end{equation}}
\newcommand{\bea}{\begin{eqnarray}}
\newcommand{\eea}{\end{eqnarray}}

\newcommand{\lbl}[1]{\label{eq:#1}}
\newcommand{\rf}[1]{(\ref{eq:#1})}
\newcommand{\nn}{\nonumber}
\newtheorem{Theorem:Newlander-Niremberger}{Theorem:Newlander-Niremberger}[section]

\newcommand{\beq}{\begin{equation}}
\newcommand{\eeq}{\end{equation}}

\begin{document}

\pagestyle{empty}

%----------------------------------------------------------------

\font\fifteen=cmbx10 at 15pt
\font\twelve=cmbx10 at 12pt

\begin{titlepage}

\begin{center}
\renewcommand{\thefootnote}{\fnsymbol{footnote}}

{\twelve Centre de Physique Th\'eorique\footnote{
Unit\'e Propre de Recherche 7061, et FRUMAM/F\'ed\'eration de Recherches 2291
}, CNRS Luminy, Case 907}

{\twelve F-13288 Marseille -- Cedex 9}

\vskip 2cm

{\fifteen  {\LARGE $\cW_\infty$}--algebras in {\LARGE $n$} complex
dimensions and\\[2mm]
Kodaira-Spencer deformations : a symplectic approach}

\vskip 1.5cm

\setcounter{footnote}{0}
\renewcommand{\thefootnote}{\arabic{footnote}}

{\bf G. BANDELLONI} $^a$\footnote{e-mail : {\tt beppe@genova.infn.it}}
\hskip .2mm  and \hskip .7mm {\bf S. LAZZARINI} $^b$\footnote{and also
Universit\'e de la M\'editerran\'ee, Aix-Marseille II. e-mail : {\tt
sel@cpt.univ-mrs.fr} }\\[6mm]
$^a$ \textit{Dipartimento di Fisica
dell'Universit\`a di Genova,}\\
{\it Via Dodecaneso 33, I-16146 GENOVA, Italy}\\
 and \\
{\it Istituto Nazionale di Fisica Nucleare, INFN, Sezione di Genova}\\
{\it via Dodecaneso 33, I-16146 GENOVA, Italy}\\[4mm]
$^b$ {\it Centre de Physique Th\'eorique, CNRS Luminy, Case 907,}\\
{\it F-13288 MARSEILLE Cedex, France}
\end{center}

\vskip 1.cm

\centerline{\bf Abstract}
It is shown that the notion of $\cW_\infty$-algebra originally carried
out over a (compact) Riemann surface can be extended to $n$ complex
dimensional (compact) manifolds within a symplectic geometrical setup.
 The relationships with the
Kodaira-Spencer deformation theory of complex structures are discussed.
Subsequently, some field theoretical aspects at the classical level are briefly
underlined.

\vskip 1.5cm

\noindent 1998 PACS Classification: 11.10.Gh - 11.25 hf - 03.70

\noindent Keywords:  Complex manifolds, Kodaira-Spencer deformation,
symplectic geometry, $W$-algebras.

\indent

\noindent{CPT--2001/P.4229}, \hfill{Internet : {\tt www.cpt.univ-mrs.fr}}

\end{titlepage}

\renewcommand{\thefootnote}{\arabic{footnote}}
\setcounter{footnote}{0}
\pagestyle{plain}
\setcounter{page}{1}

\newpage

\sect{Introduction}

It is fair to say that the concept of dimensionality plays an
important role in Physics. In particular,
the developments in quantum field theory as well as in statistical mechanics
have greatly enlarged its importance.
In renormalization theory, string field models,
the concept of dimension is found to be
not only a characterization of the background space were the physical
phenomena are supposed to take place, but also a
physical regularizing parameter.
Indeed, a world with a given dimension very often shows
merits and faults
not  found in some others of different dimensions.
This led to the search for hyperspaces which could gather together the
praises and avoid the imperfections of the theoretical models.

For instance, two dimensional models show the great relevance of
 complex structures \cite{BK} in Quantum Field Theory.  Moreover, this
 approach produces a dimensional halving, but, in spite of the low
 dimensionality, the conformal models are described by  means of an
 infinite dimensional algebra \cite{BPZ}.

So, the wide class of these ``new" symmetries has been supporting the
 conjecture that life in two dimensions could be easier and more
 convenient \cite{Dew}. The so-called $\cW$-algebras \cite{Zam} were a
 byproduct of this feasibility in two dimensional spaces. For an
 extensive review on the various possibilities offered by these kinds
 of symmetries we refer to \cite{Tjin}.
Thus the question of extending this type of symmetries to
 higher dimensional spaces comes naturally.
The extension required the use of the Kodaira-Spencer deformation
 theory \cite{Kod86}.
In particular, chiral symmetries have already been extended from $2D$ conformal
models built on  a Riemann surface to models to a $n$ complex
dimensional complex manifold \cite{BCOV94,LMNS97,BL98}. Note that
Kodaira-Spencer type deformation theories have been already used to describe
$\cW_\infty$ in two (or more)  dimensions
\cite{Fairlie:1990wv,Dijkgraaf:1997iy,Castro} in order to study holomorphic
properties (chiral splitting) or mirror manifolds of arbitrary complex
 dimension \cite{BCOV94,LMNS97,Vaf94,BS94,Barannikov,LM94}. Their cohomologies
have been investigated both in Lagrangian Field Theory models  \cite{BL98}
and in more general mathematical aspects in \cite{Kos60,Fucks86}.

Therefore we shall address in the present paper the extension to $n$
complex dimensions of our BRS treatment for $\cW_\infty$-algebra
grounded on a symplectic approach 
\cite{symplecto,complexstruct,w3}. In the latter, the algebra emerges
from a ghost realization geometrically constructed from
the symplectic approach and as a byproduct the infinite number of
chiral ghost 
fields $\cC^{(n)}$, $n=1,2,...$, turn out to be
$(-n,0)$-conformal fields and their infinitesimal variations have a
well defined geometrical setting.  
To be more specific, let us remind how the chiral $\cW_\infty$-algebra is
recovered in the 
bidimensional case over a Riemann surface. For any positive integer
$n$, the local variations of the chiral ghosts are 
\begin{eqnarray}
\cS \cC^{(n)}\zbz = \sum_{m=1}^n m\, \cC^{(m)}\zbz \prt_z
\cC^{(n-m+1)}\zbz.
\lbl{n1}
\end{eqnarray}
Introducing by duality to each ghost a local operator $T_{(n)}\zbz$
in order to construct the anticommuting functional BRS operator
\cite{Becchi:1975nq},
\begin{eqnarray}
\delta = \sum_{n\geq 1} \int_{\Sigma}
d\bz\wedge dz \biggl(\cC^{(n)}\zbz T_{(n)}\zbz + \cS \cC^{(n)}\zbz
\frac{\delta}{\delta \cC^{(n)\zbz}} \biggr),
\lbl{BRSfunct}
\end{eqnarray}
namely, $\{\delta,\delta\} = 0$, leads to the following local
commutation relations,
\begin{eqnarray}
\biggl[ T_{(n)}\zbz, T_{(m)}\zbzp \biggr]
= n\, \prt_{z'}\delta^{(2)}(z'-z) T_{(n+m-1)}\zbz 
- m\, \prt_z \delta^{(2)}(z-z') T_{(n+m-1)}\zbzp 
\lbl{Walg}
\end{eqnarray}
which turn out to be a realization of the so-called
$W_{\infty}$-algebra if one goes to the Fourier modes.

We stress that the well defined ghost realization allows one to write down the
extension of the $\cW_\infty$-algebra to higher dimensions.  
Moreover we want to take advantage of the symplectic description for
 incompressible flows in order to extend to $n$ dimensions the notion of
$\cW_\infty$-algebra, which in two dimensions is related to
area preserving diffeomorphisms, see for instance \cite{She92} and
references therein. 

The algebra will be
 described in our approach by means of the Kodaira-Spencer deformation
 theory of complex structures but reformulated in a symplectic
 framework.  The physical motivation of investigating  the subject
 relies is connected to the so-called $W$-gravity and also 
on the fact that quantizing  a conformal gravitational theory
 would incorporate all  the possible configurations of the
 gravitational fields.  By the way, ``well defined" gravitational
 conformal models are fully described by means of the complex
 structure of the surrounding space. Therefore a complete description
 just at the classical level of all its possible deformations might be
 relevant  for a successful quantum improvement.  

The paper is
 organized as follows.  We shall first briefly introduce in a
 non-technical way, the Kodaira-Spencer deformations, referring the
 reader to the book by Kodaira \cite{Kod86} for a more complete
 survey, especially Chapters 2, 4 and 5.  Then Section 3 will  give a
 geometrical setting of symplectomorphisms in a generic $n$ complex
 dimensional space in order to introduce the BRS formulation of the
 (infinitesimal) diffeomorphisms of a symplectic space.  Furthermore
 in Section 4 the specific Kodaira-Spencer deformation of complex
 structures related to $\cW_\infty$-algebra will be presented through
 a symplectic approach by using a ghost 
representation.  It is recalled that the symplectic treatment of the
 two-dimensional case for $\cW$-algebras \cite{symplecto} provides
 a well defined geometrical definition of the ghost fields and
 their BRS variations as well. In the present paper we avail
 ourselves of that symplectic approach, in order to address
 the problem of 
 extending to arbitrary complex dimensions the notion of
 $\cW_\infty$-algebra and its consequences, in particular, for  the
 study of Lagrangians subject to that type of symmetry to which a very
 brief Section will be devoted.

\section{A short account on the Kodaira-Spencer deformation}

Let $M$ be a $n$ dimensional (compact) complex manifold described in
terms of background local complex coordinates:
\be
(z^k) :=(z^1,z^2,\cdots, z^n), \qquad k=1\cdots n
\ee
and the subordinated differentiable structure $\zbzk$ turns $M$ into a
$2n$ real dimensional manifold.

Its complex structure is determined by the ${\displaystyle
\bprt\equiv\sum_{i=1}^nd\bz^{\bi} \bprt_{\bi}}$ operator. In order
to control the deformation, usually a complex deformation
parameter $t=(t_1,\dots,t_n) $ is introduced.  Basically the
physical implications of this mathematical field of interest,
rises from the primitive idea that a complex manifold is composed
of a set of coordinate neighborhoods patched together. Obviously
the patching procedure sewing  should be irrelevant  to the
manifold description. In this philosophy a deformation of $M$ is
considered to be the sewing  of the same patches, through a fit of
the parameters $t$
 via various identifications.
 Four our purpose the dimension of
the parameter space will be exactly equal to that of $M$.
According to Chapter 5 of \cite{Kod86} one considers a complex family
of compact complex
manifolds as a complex manifold $\cM$ and a holomorphic
map $\varpi:\cM\rightarrow \cB$ where $\cB$ is a domain in ${\mathbb
C}^n$ such that $\varpi^{-1}(t)=M_t$ is a compact complex
manifold. For $\Delta\subset\cB$ sufficiently small, $\cM_{\Delta}
:= \varpi^{-1}(\Delta)$ can be identified as a complex manifold with
the complex structure defined on the smooth manifold $M\times\Delta$
since the subordinated smooth structure is always the same and does
not depend on $t$ (\cite{Kod86} Thm 2.3). Accordingly, local complex
coordinates on
$\cM_{\Delta}$ will be given by the system of local complex
coordinates $(\cZ^\a\zt,t^\a)$, $\a=1,\dots,n$ and
for fixed $t$, $M_t$ is the complex structure of the differentiable
manifold $M$ defined by the system of local complex coordinates
$(\cZ^\a\zt)$, $\a=1,\dots,n$  considered as a smooth change of local
complex coordinates on $M$ i.e. the Jacobian does not vanish.

On the other hand, the deformation of complex structure is thus
described by the change of the $\bprt$-operator \cite{Kod86}
\be
\bprt\lra \bprt-\sum_{\ell=1}^n \mu^\ell\zt\prt_\ell\ ,
\ee
the $\mu^\ell\zt$ are unique smooth $(0,1)$-forms on
$M\times\Delta$.
In this way, one can describe both infinitesimal and finite deformations.
Indeed, by looking for, at fixed $t$, the local solutions
$\cZ^\a\zt$ of this family of deformed $\bprt$-operators
\bea
&&\Big(\bprt- \sum_{\ell=1}^n \mu^\ell\zt\prt_\ell\Big)\cZ^\a\zt=0
\lbl{kod}
\eea
then they will patch together holomorphically with respect to the
complex structure $M_t$ and thus they will define a new
complex structure parametrized by the $\mu$ on $M$.

To be consistent { with the deformation philosophy discussed
before} , the previous equation \rf{kod} must be coupled
(Newlander-Nirenberg integrability theorem) \cite{Kod86} with the
Kodaira-Spencer (integrability) equation \be
\bprt\mu\zt-\frac{1}{2}\Big[\mu\zt,\mu\zt\Big]=0 \lbl{kod1} \ee
where $\mu\zt =\mu^\ell\zt\prt_\ell$, is a smooth $(1,0)$-vector
field valued $(0,1)$-form on $M$ and the graded brackets $[,]$
means the commutator of two vector fields and wedging.

To sum up, two solutions of Eq\rf{kod1} correspond to the same complex
 structure if they differ by an holomorphic diffeomorphism. Since for
 $t=0$ both $(z^k)$ and $(\cZ^\a(z,0))$ are local complex coordinates
 on the complex manifold $M$, then $\cZ^\a(z,0)$ are holomorphic
 functions of $(z^k)$, showing that $\mu(z,0)=0$. The construction of
 the new local complex coordinates $\cZ^\a(\zbz,t)$ for each fixed $t$ will
 correspond to a smooth change of local complex coordinates $(z^k) \mapsto
(\cZ^\a(\zbz,t))$. The construction holds in
each holomorphic sector in $t$. Embedding in a symplectic
 framework  generates an infinite sequence of changes of local complex
 coordinates. It is the signature of their behavior under
 symplectomorphisms which gives rise 
 to an algebra. The latter extends to higher dimensions the usual
 $\cW_\infty$-algebra \rf{Walg}. 

For this reason, if we wish now to settle the Kodaira-Spencer
deformation in a symplectic framework, we may consider the
deformation parameters as the conjugate variables (by
symplectic doubling, as it will be better specified later on) to
 those of the configuration space by identifying locally, as
differentiable manifolds, the
cotangent space $T^*M$ with $\cM_{\Delta}$ endowed with local smooth
coordinates $(z^k,\bz^\bk,t,\ovl{t})$. Then 
all the requirements to perform a Kodaira-Spencer deformation will
be satisfied, so that this mathematical artillery will be at our
disposal to investigate the possible extension of our symplectic
approach to a $n$ complex dimensional manifold and the
consequences for physical models, in particular higher spin fields
and their sources.

\sect{Symplectomorphisms in $2n$ complex dimensional complex symplectic space}

Symplectomorphisms describe diffeomorphisms preserving a given
symplectic structure on the cotangent bundle $T^*M$. They can be
respectively described in terms of local coordinates, namely,

\bea
&&\cU(z,y)=(z^1\cdots z^n, \bz^1 \cdots \bz^n; y_1 \cdots y_n ,\by_1
\cdots \by_n),\nn\\
\lbl{phasespacezy}
&&\cU(Z,\cY)=(Z^1\cdots Z^n, \bZ^1 \cdots \bZ^n; \cY_1 \cdots \cY_n ,\bcY_1 \cdots \bcY_n)
\lbl{phasespaceZY}
\eea
and respectively endowed with the symplectic fundamental 2-form which,
in full generality, locally writes according to the  system of local
coordinates -not necessarily the Darboux's ones,

\bea
\Omega_{\cU(z,y)}= \sum_{i,j=1}^n\omega_j^i dz^j\wedge dy_i +
c.c. =d\theta_{\cU(z,y)}
\lbl{omegazy}\\
\Omega_{\cU(Z,\cY)}=\sum_{\a,\b=1}^n \omega_\b^\a dZ^\b\wedge
 d\cY_\a + c.c.  =d\theta_{\cU(Z,\cY)}
\lbl{omegaZY}
\eea
with the following local requirements
\bea
\det{|\omega_{(i,j)}|}\neq 0;\quad
\det{|\omega_{(\b,\a)}|}\neq 0;\quad
d_z\omega_j^i=d_y \omega_j^i=d_Z\omega_\b^\a=d_\cY\omega_\b^\a =0.
\lbl{condomega}
\eea

and the invariance of the fundamental 2-form is locally expressed
by \bea &&\Omega_{\cU(z,y)}=\Omega_{\cU(Z,\cY)}. \lbl{canonical}
\eea Locally, this implies on
${\cU(z,y)}\bigcap{\cU(Z,\cY)}\neq\emptyset $ that \bea
&&\theta_{\cU(z,y)}-\theta_{\cU(Z,\cY)}=dF. \lbl{dFl} \eea

From now on, we shall work locally in terms of the `mixed' local
 independent coordinates $\zY$,

  \bea
&&\cU(z,\cY)=(z^1\cdots z^n, \bz^1 \cdots \bz^n; \cY_1 \cdots \cY_n ,\bcY_1 \cdots \bcY_n)
\lbl{phasespacezY}
\eea

where we define the differential operators (from now on the Einstein's
convention for summation will be used throughout the paper):

\bea
d=d_z +d_\cY;\quad
d_z= dz^i\frac{\prt}{\prt z^i} +d\bz^{\bi} \frac{\prt}{\prt \bz^{\bi}}
=dz^i\prt_i + d\bz^{\bi}\prt_{\bi};\quad
d_\cY=  d\cY_\a \frac{\prt}{\prt \cY_\a} +d\bcY_{\ba} \frac{\prt}{\prt \bcY_\a}
\lbl{differentialss}
\eea

The corresponding generating function $\Phi\zY$ is obtained
trough the Legendre transformation

\bea
d\Phi(z\cY) = d(F + \omega_\b^\a Z^\b\cY_\a+c.c.) =
\omega_j^i y_i dz^j + \omega_\b^\a Z^\b d\cY_\a + c.c.
\lbl{genfun}
\eea
In the cotangent space $T^*M$ endowed with this system of local
coordinates, the mappings: 
\bea
\mbox{\boldmath $y$}_i\zY\equiv \omega^j_i y_j&=&\frac{\prt\Phi\zY}{\prt z^i}\equiv \prt_i\Phi\zY \lbl{mappcanonical0}\\
\cZ^\a\zY\equiv \omega^\a_\b Z^\b &=& \frac{\prt\Phi\zY}{\prt
\cY_\a} \lbl{mappcanonical} \eea are canonical and define new
canonical variables via the $\omega$ matrices.

Several ways can settle this canonical procedure: Poisson brackets (or
something similar), flow analysis of hierarchical structures.
 We shall be concerned with the study of this aspect in a field
theoretical language by using the BRS formulation.
Moreover we can rewrite:

\bea
\Omega_{\cU(z,\cY)}& =&\prt_i\cZ^\a\zY dz^i\wedge d\cY_\a +\bprt_{\bi} \cZ^\a\zY d\bz^{\bi}\wedge d\cY_\a\nn\\
&+&\prt_i\bcZ^{\ba}\zY dz^i\wedge d\bcY_{\ba}+\bprt_{\bi} \bcZ^{\ba}\zY d\bZ^{\ba}\wedge d\bcY_{\ba}   \nn     \\
&=& \frac{\prt}{\prt \cY_\a}\bby_i\zY dz^i\wedge d\cY_\a + \frac{\prt}{\prt \cY_\a}\obby_{\bi}\zY d\bz^{\bi}\wedge d\cY_\a\nn\\
&+& \frac{\prt}{\prt \bcY_{\ba}}\bby_i\zY dz^i\wedge d\bcY_{\ba}+ \frac{\prt}{\prt \bcY_{\ba}}\obby_{\bi}\zY d\bz^{\bi}\wedge d\bcY_{\ba}\nn\\
&=&d_z d_\cY\Phi(z,\cY)
\lbl{omegazY}
\eea
from which we get the relations of duality (with their complex
conjugate expressions as well):
\be
\prt_i\cZ^\a\zY= \frac{\prt}{\prt \cY_\a}\bby_i\zY ,
\qquad
\bprt_{\bi} \cZ^\a\zY=\frac{\prt}{\prt \cY_\a}\obby_{\bi}\zY  .
\lbl{duality}
\ee

In order to parametrize our space we define
 \cite{symplecto,complexstruct,w3} the Hessian matrix elements by
 \bea
&&\prt_i\frac{\prt}{\prt\cY_\a}\Phi\zY\equiv \lambda^\a_i\zY
\lbl{lambda} \\[2mm]
&&\bprt_{\bj}\frac{\prt}{\prt\cY_\a}\Phi\zY\equiv \lambda^\a_i\zY\mu^i_{\bj}\zY\equiv \blambda^{\bb}_{\bj}\zY\mub_{\bb}^{\a}\zY
\lbl{lambdamu}
\eea
with $\det|\lambda|\neq 0$ for non singularity requirement and also
 for the complex conjugate expressions.
 From Eqs\rf{lambda}\rf{lambdamu} we get the following identities:
\bea \prt_j \lambda^\a_i\zY=\prt_i \lambda^\a_j\zY &;&
\frac{\prt}{\prt\cY_\b}\lambda^\a_i\zY=
\frac{\prt}{\prt\cY_\a}\lambda^\b_i\zY
\lbl{lambda1}\\[2mm]
\bprt_{\bj}\lambda^\a_i\zY=\prt_i\biggl(\lambda^\a_r\zY\mu^r_{\bj}\zY\biggr)
&;&
\frac{\prt}{\prt\bcY_{\bb}} \lambda^\a_i\zY
=\frac{\prt}{\prt\cY_\a}\biggl(\blambda^{\bb}_{\br}\zY\mub^{\br}_i\zY\biggr)
\lbl{beltrami1}\\[2mm]
\bprt_{\bj}\lambda^\a_i\zY=\prt_i\biggl(\blambda^{\ba}_{\bj}\zY\mub^\a_{\ba}\zY \biggr)
&;&\frac{\prt}{\prt \bcY_{\bb}}\lambda^\a_i\zY =\frac{\prt}{\prt \cY_\a}
\biggl( \lambda^\sigma_i\zY\mu^{\bb}_{\sigma}\zY \biggr)
 \lbl{beltrami2}
\eea

So from Eqs \rf{duality}\rf{lambda}\rf{lambdamu} we have the
following two main identities which must be viewed within the  Kodaira-Spencer
spirit of Eq\rf{kod} :
\bea
\biggl(\bprt_{\bj}- \mu^r_{\bj}\zY\prt_r\biggl)\cZ^\a\zY
\equiv \cL_{\bj}\zY\cZ^\a\zY=0
\lbl{Z}
\eea
\bea
 \biggl(\frac{\prt}{\prt\bcY_{\ba}}
-\mu^{\ba}_{\b}\zY\frac{\prt}{\prt\cY_\b}\biggr)\bby_r\zY\equiv
\cL^{{\ba}}\zY\bby_r\zY =0 \lbl{y} \eea where the role of the
parameter $t$ of deformation is presently played by the covariant
coordinates $(\cY,\bar{\cY})$ in the former or by the background
complex coordinates $\zbz$ in the latter. The first of the two
equations tells that a local deformation of the complex structure
on the base complex manifold $M$ can be implemented by using the
symplectic structure on the cotangent bundle $T^*M$, while the
second one governs the vertical deformation. This coincidence
justifies our point of view of taking the conjugate variables as
the deformation parameter. Hence, the complex family of complex
manifolds $M_\cY$ is locally recasted as the symplectic cotangent
bundle $TM$ when the differentiable structure is considered.

Let us write down the following Pfaff system

\bea
d_z\cZ^\a\zY&=&
\lambda^\a_i\zY\biggl(dz^i+\mu^i_{\bj}\zY d\bz^{\bj}\biggr) =:
\Big(dz + d\bz\pt\mu\zY \Big)\pt \lambda\zY
\nn\\
d_\cY\bby_i\zY
&=& \lambda^\a_i\zY \biggl( d\cY_\a +\mu^{\ba}_\a\zY
d\bcY_{\ba}\biggr)
\lbl{differentials}
\eea
The system serves to define two types of Kodaira-Spencer differentials, namely,
$\mu^i_{\bj}\zY $ and
$\mu_{\ovl{\b}}^\a\zY $ which parametrize the complex structures on
the base space $M$ with background local complex coordinates $\zbz$
and the fibers with local coordinates
$\YBY$, respectively. These complex structures are interlinked by the
duality relations Eqs\rf{duality}\rf{lambdamu}
\bea
\mu^i_{\bj}\zY
&=& \bprt_{\bj} \cZ^\b\zY\, [\lambda\zY^{-1}]_\b^i
\lbl{muij}\\
\mu_{\b}^{\ovl{\a}}\zY
&=& [\lamb\zY^{-1}]_\b^r\,\prt_r\ovl{\cZ}^{\ovl{\a}}\zY.
\lbl{muab}
\eea
Inverting the previous formulas Eq\rf{differentials} by matrix inversion

\be
\prt_i  = \lam^\a_i\zY\prt_\a + \mub^\jb_i\zY \lamb^{\ovl{\a}}_
\jb \zY\prt_{\ovl{\a}}
= \lam^\a_i\zY\Big(\prt_\a + \mu^{\ba}_\a\zY \prt_{\ovl{\a}}\Big)
\ee

where $\prt_\a={\displaystyle \frac{\prt}{\prt\cZ^\a\zY}} $, one gets

\be
\frac{\prt}{\prt\cY_\a} = \lam^\a_i\zY\biggl(\frac{\prt}{\prt
\bby_{i}} +\mu^i_\jb\zY  \frac{\prt}{\prt
\obby_\jb}\biggr) \equiv \lam^\a_i\zY\cD^i\zY .
\lbl{derY}
\ee

It now easy to derive from Eqs\rf{lambdamu}\rf{derY} another
description of the Kodaira-Spencer differentials
\be
\mu^i_{\bj}\zY=\cD^i\prt_{\bj}\Phi\zY,
\lbl{muPhi}
\ee
where $\cY=\cY(z,y)$ has to be taken into account.
The most relevant properties of the $\cD^i\zY$ and $ \frac{\prt}{\prt
\bby_{i}} $ operators can be summarized as

\bea
\biggl[\cD^i\zY,\cD^j\zY\biggr]=0,\quad
\biggl[ \frac{\prt}{\prt \bby_{i}}, \frac{\prt}{\prt \bby_{j}}\biggr]=0,\quad
\biggl[ \frac{\prt}{\prt \bby_{i}}, \frac{\prt}{\prt
\obby_{\bi}}\biggr]=0.
\lbl{yby}
\eea
The third order derivatives of $\Phi$ yields
The integrability conditions Eq\rf{kod1} for the deformation of complex
structures in the $\zbz$ and $\YBY$ spaces respectively write

\bea
\cL_{\bi}\zY\mu^s_{\bj}\zY=\cL_{\bj}\zY\mu^s_{\bi}\zY
\lbl{ks}
\eea
\bea
\lambda^{\b}_i\zY\prt_j\mu^{\ba}_\b\zY = \lambda^\b_j\zY\prt_i\mu^{\ba}_\b\zY
\lbl{compatibility}
\eea

Moreover in the $\YBY$ space, the partner of the Kodaira-Spencer
 equations can be immediately recovered computing $\prt_{j} \frac{\prt}{\prt\bcY_{\ba}}\frac{\prt}{\prt \bcY_{\bb}}\Phi\zY $
\bea
\cL^{\ba}\zY\mu^{\bb}_\lambda\zY =\cL^{\bb}\zY\mu^{\ba}_\lambda\zY
\lbl{integrability}
\eea

with the consistency conditions:
\bea
\biggl[\cL_{\bi}\zY ,\cL_{\bj}\zY\biggr]=0
&;\quad&
\biggl[\cL^{\ba}\zY ,\cL^{\bb}\zY\biggr]=0 .
\lbl{commutations}
\eea

\subsection{BRS setting of symplectomorphisms in $2n$ complex dimensions }

As said before the Kodaira-Spencer deformations reparametrize in a
consistent way  the space of complex structures.  Furthermore,
we shall study the action of reparametrizations on symplectic space
(symplectomorphisms).

The BRS setting for   symplectomorphisms can be performed
along the lines developed in \cite{symplecto}.
Let us define by  $\cS$ the nilpotent BRS operation associated to the
infinitesimal 
symplectomorphisms. Locally, $\cS$ will be represented in $\zY$
coordinates by
\be
\cS \Phi\zY=\Lambda\zY, \qquad \cS\Lambda\zY = 0
\ee
The infinitesimal BRS transformation of the deformed coordinate
 $\cZ^\a\zbz$ can be calculated from its canonical definition
 Eq\rf{mappcanonical0},\rf{mappcanonical}
\be
\cS \cZ^\a\zY=\frac{\prt}{\prt \cY_\a}\Lambda\zY=
 \lambda^\a_i\zY\cD^i\zY \Lambda\zY = \cC^i\zY\prt_i\cZ^\a\zY
\lbl{SZ}
\ee
where the chiral ghost fields $\cC^i\zY$  naturally emerge and
are related to the ordinary
 diffeomorphism ghosts $c^i\zY$,
 $\bar{c}^{\bj}\zY$ on $T^*M$ within this symplectic framework by
\bea
\cC^i\zY\equiv\cD^i\Lambda\zY= \left(
\frac{\prt\Lambda\zY}{\prt \bby_{i}} +\mu^i_\jb\zY
 \frac{\prt\Lambda\zY}{\prt \obby_\jb} \right)
= c^i\zY + \mu^i_{\bj}\zY \bar{c}^{\bj}\zY
\eea
which explicitly corresponds to a change of generators for symplectomorphisms.
Their BRS variations read
 \bea
 \cS\cC^i\zY &=& \cC^j\zY\prt_j \cC^i\zY\\
 \cS c^i\zY &=&  \Big[c^j\zY\prt_j+ c^{\bj}\zY \bprt_{\bj}\Big]c^i\zY .
 \eea
 These BRS transformations correspond to an
  infinitesimal reparametrization
   of $\cZ^\a\zY$ due to an infinitesimal shift of the $\zbz$
 background, keeping $\YBY$ fixed.

We can easily derive:
\bea
\cS\lambda^\a_i\zY=\prt_i\Big( \lambda^\a_j\zY\cC^j\zY\Big)\\
\cS\Big(\lambda^\a_r\zY \mu^r_{\bj}\zY\Big) =
\bprt_{\bj}\Big( \lambda^\a_r\zY\cC^r\zY\Big)
\eea
so that

\be
\cS\mu^i_{\bj}\zY=\cC^l\zY\prt_l\mu^i_{\bj}\zY -\prt_l\cC^i\zY\mu^l_{\bj}\zY +\bprt_{\bj}\cC^i\zY
\ee

The non-chiral representation of this algebra can be easily given following the lines of \cite{BL98}, where we have stressed the relevance
of the $\zbz $ counterpart of  the Kodaira-Spencer equation \rf{integrability}.

Moreover the ordinary ghosts  $c^i\zY$ transform as:
\be
\cS c^i\zY = \Big( c^l\zY\prt_l  +\bc^{\bl}\zY\bprt_{\bl}\Big)   c^i\zY
\ee

Finally, note the important commutators coming from the combination of
the commutators ${\displaystyle
[\cS,\frac{\prt}{\prt\cY_{\a}}]=0=[\prt_i,\frac{\prt}{\prt\cY_{\a}}]}$
with \rf{lambda1} \rf{derY}:
\bea
\Big[\cS,\cD^i\zY\Big] &=& -
[\lambda\zY^{-1}]_\a^i\, \prt_i\Big( \lambda^\a_j\zY\cC^j\zY\Big)\,
\cD^j\zY \nn\\
&{}&\lbl{sd}\\[-3mm]
&=& - \prt_r\cC^i\zY \cD^r\zY + \cC^r\zY \Big[\prt_r,\cD^i\zY\Big]. \nn
\eea

Conversely,  from Eqs\rf{mappcanonical0},\rf{mappcanonical} we can
derive
 the infinitesimal transformation of $\bby_{i}\zY $ due to an infinitesimal reparametrization
 on $\YBY$ space, keeping the $\zbz$ background fixed.

\bea
\cS \bby_{i}\zY=\prt_i\Lambda\zY& =& \lam^\a_i\zY\Big(\prt_\a +
\mu^{\ba}_\a\zY \prt_{\ovl{\a}}\Big) \Lambda\zY\nn\\
               &=& \Big(\omega_\a\zY + \mu^{\ba}_\a\zY
\omega_{\ovl{\a}}\zY \Big) \frac{\prt}{\prt \cY_\a}\bby_i\zY \nn\\
                &=& \cO_\a\zY \frac{\prt}{\prt \cY_\a}\bby_i\zY
\eea
where it has been set
\be
\omega_\a\zY=\prt_\a\Lambda\zY
\ee
and:
\bea
\cS\cO_\a\zY = \cO_\b\zY\frac{\prt}{\prt \cY_\b} \cO_\a\zY.
\eea

Now the generating function $\Phi\zY$ for such canonical
transformations will be so chosen in order to view the holomorphic
deformation process in the $\cY$ direction 
as being a canonical transformation.

For the purpose it will be convenient to use a multi-index notation.
Let $A,B$ denote multi-indices on the fibers related to
Greek indices and while and $I,J$ denote multi-indices on $M$ related
to Latin indices.
For $A=(a_1,\dots,a_n)$ with positive integers $a_\a\geq 0$, $|A| =
{\displaystyle \sum_{\a=1}^n a_k}$ will be the order of $A$ and one sets
$A+1_\b = (a_1,\dots,a_{\b-1},a_\b+1,a_{\b+1},\dots,a_n)$,
$A! = {\displaystyle \prod_{\a=1}^n a_\a!}$.
For the sake of notational completeness, on the base $M$ one will
similarly use $I+1_k = (i_1,\dots,i_{k-1},i_k+1,i_{k+1},\dots,i_n)$.
Now, one {\em chooses} a $\cY$-holomorphically split generating function
\be
\Phi\zY= \sum_{|A|\geq 1} \cZ^{(A)}\zbz\, \cY_A + c.c.,
\lbl{holphi}
\ee
where for $|A|\geq 1$ and ${\displaystyle \cY_A := \prod_{\a=1}^n
(\cY_\a)^{a_\a} }$, we have set
\begin{eqnarray}
\cZ^{(A)}\zbz :=
\frac{1}{|A|!}\frac{\prt^{|A|}\Phi\zY}{\prt
\cY_A}\Big|_{\cY=0} :=
\frac{1}{|A|!}\frac{\prt^{|A|}\Phi\zY}{(\prt
\cY_1)^{a_1}\cdots(\prt\cY_n)^{a_n}
}\Big|_{\cY=0}
\lbl{jetZ}
\end{eqnarray}
for the ${\displaystyle \frac{(|A|+n-1)!}{|A|!\,(n-1)!} }$
independent derivatives of order $|A|$. With such a generating
function the symplectic two-form \rf{omegazY} is locally written as 
\bea
\Omega = \sum_{|A|\geq 1} d_z \cZ^{(A)}\zbz \wedge d_\cY \cY_A +
c.c \lbl{Omegaholo} \eea 
while the new coordinates defined in \rf{mappcanonical0} and
\rf{mappcanonical} are respectively given by 
\begin{eqnarray}
\bby_{i} \zY &=& \sum_{|A|\geq 0}\ \prt_i\cZ^{(A)}\zbz\,\cY_A +
\sum_{|\ovl{B}|\geq 0}\ \prt_i \ovl{\cZ}^{(\ovl{B})}\zbz\,
\ovl{\cY}_{\ovl{B}}\,. \lbl{yY}\\
\cZ^\a\zY &=& \sum_{|A|\geq 0}\ \sum_{\a=1}^n (a_\a +1)
\cZ^{(A+1_\a)}\zbz\,\cY_A\nn\\
&=& \cZ^\a\zbz + \sum_{|A|\geq 1}\ \sum_{\a=1}^n
(a_\a +1)
\cZ^{(A+1_\a)}\zbz\,\cY_A
\lbl{ZY}
\end{eqnarray}
Note that $\cZ^\a(z,\cY)\Big|_{\cY=0} = \cZ^\a\zbz$
showing that the complex structure given by the local complex
coordinates $\cZ^\a$ is the one which is actually deformed.
Recall that the latter are local complex coordinates
solutions of \rf{Z} at $\cY_\a=\bcY_{\ba}=0$ and have already been
treated in the context $n$ complex dimensional manifolds in \cite{BL98}.

As explicitly shown above, the local coefficients
$\cZ^{(A)}\zbz$, $|A|\geq 1$ thus describe the response to the
deformation of the $\cZ^\a\zbz$ complex coordinates. 
Combining the decomposition \rf{Omegaholo} with the covariance
requirement \rf{canonical} leads to an infinite sequence of
 changes of local complex coordinates
 $(z^k)\lra (\cZ^{(A)}\zbz)$ whose the algebra of infinitesimal
transformations can be derived by means of BRS techniques.

Furthermore, the role of
the complex structures involved in the present approach can be deepened.
Indeed, the Kodaira-Spencer differentials
$\mu^i_{\bj}\zY$ reflect the general behavior
(see Eq \rf{muPhi}) of the generating function of the canonical
transformations.
Their infinitesimal behavior in the $\zbz$ and $\YBY$
spaces are constrained by both Eq \rf{compatibility} and \rf{integrability}.
Now the explicit complex deformation will be chosen
as a particular case of \cite{Kod86}, according to
\begin{eqnarray}
\mu_{\bj}^i\zY &=& \sum_{|A|\geq 0} \ \mu_{\bj}^i\, ^{(A)}\zbz\, \cY_A\
,\lbl{mudeform}\\
\mbox{with }\
\mu_{\bj}^i\, ^{(A)}\zbz &=& \frac{1}{|A|!}\frac{\prt^{|A|+1}}{\prt
\cY_{A+1_\b}}\, \Big(\bprt_{\bj} \Phi\zY \,
[\lambda\zY^{-1}]_\b^i\Big)\Big|_{\cY=0}  .
\nn
\end{eqnarray}
This series converges in a Holder norm \cite{Kod86} and represents
a deformation of the integrable complex structure defined by
$\mu_\bj^i{}^{(0)}$ with the role of deformation parameters is
played by $\cY$ as already said before. Since the use of this
space doubling is to introduce a symplectic structure in order
that the smooth local changes of complex coordinates $(z^k)\lra
(\cZ^{(A)}\zbz)$ are interpreted as coming from a
symplectomorphism symmetry.
 Recall that the generating function
\rf{holphi} for the canonical transformations
has been chosen to be compatible with the deformation
\rf{mudeform}.
The holomorphic character of the deformations  will define,
 in a BRS framework, a series of infinitesimal symmetry transformations
 which will reproduce the $n$ complex dimensional extension of the
$\cW_\infty$-algebra as will be shown in the next Section.

The link of the parametrization in Eq\rf{ZY} with the one of
\rf{mudeform} is given through \rf{lambdamu} by, for $|A|\geq 0$ and
for each $\a=1,\dots,n$ --no summation over $\a$--
\begin{eqnarray}
(a_\a+1) \bprt_{\bj}\cZ^{(A+1_\a)}\zbz = \sum_{\scriptsize \begin{array}{c}
|B|,|C|\geq 0\\ B+C=A \end{array} }\
(b_\a+1) \prt_i \cZ^{(B+1_\a)}\zbz\, \mu_{\bj}^i\ ^{(C)}\zbz
\lbl{link}
\end{eqnarray}
which, in the particular case of $|A|=0$, reduces to the usual
Beltrami equations
\begin{eqnarray}
\bprt_{\bj}\cZ^\a\zbz = \prt_r \cZ^{\a} \zbz\, \mu_{\bj}^r\ ^{(0)}\zbz
\lbl{2d-beltra}
\end{eqnarray}
which were fully treated in \cite{BL98} in the two dimensional case.
In this context the integrability condition \rf{ks} is transfered on
the jet coordinates $\mu_{\bj}^i\ ^{(A)}\zbz$ with $|A|\geq 0$, as follows
\begin{eqnarray}
\bprt_{\bi} \mu_{\bj}^r{}^{(A)}\zbz - \bprt_{\bj} \mu_{\bi}^r {}^{(A)}\zbz =
\hskip -5mm \sum_{\scriptsize \begin{array}{c}
|B|,|C|\geq 0\\ B+C=A \end{array} } \hskip -4mm
\Big( \mu_{\bi}^s {}^{(B)}\zbz \prt_s \mu_{\bj}^r {}^{(C)}\zbz
- \mu_{\bj}^s {}^{(B)}\zbz \prt_s \mu_{\bi}^r {}^{(C)}\zbz \Big).
\end{eqnarray}

For $|A|\geq 1$,
\begin{eqnarray}
\bprt_\bj \cZ^{(A)}\zbz = \frac{1}{A!}\, \frac{\prt^{|A|}}{\prt\cY_A}
\bprt_\bj \Phi\zY \Big|_{\cY=0} = \sum_{1\leq |I|\leq|A|}
\cG^{(A)}_{(I)}\zbz \, \mu_\bj^{(I)}\zbz
\lbl{cG}
\end{eqnarray}
where we have set for $|I|\geq 1$,
\begin{eqnarray}
\mu_\bj^{(I)}\zbz := \frac{1}{I!}\, \cD^{(I)}\zY \bprt_\bj \Phi\zY
\Big|_{\cY=0} := \prod_{k=1}^n \left( \frac{1}{i_k!}\,
\Big(\cD^k\zY\Big)^{i_k} \right) \bprt_\bj \Phi\zY\Big|_{\cY=0}
\lbl{muI}
\end{eqnarray}
as representing the $n$-dimensional version for the $\cW$-extension of the
Beltrami multipliers introduced by Bilal Fock and Kogan \cite{BFK}.
It is worthwhile to say that the coefficients $\cG^{(A)}_{(I)}\zbz$ are
very intricate non local expressions depending on the derivatives up
to order $|A|$ of $\cZ^{(B)}$, with $1\leq |B|\leq |A|$. Writing
\rf{cG} in more precise terms one has for $|A|\geq 1$,
\begin{eqnarray}
\bprt_\bj \cZ^{(A)}\zbz &=& \mu^r_\bj\zbz \prt_r\cZ^{(A)}\zbz
+ \cdots \nn\\
&& +\ \sum_{\a=1}^n \sum_{|I^\a| = a_\a}
\frac{(I^1 +\cdots +I^n)!}{I^1!\cdots I^n!} \left( \prod_{\b=1}^n
\lam^\b_{I^\b}\zbz \right) \mu^{(I^1 +\cdots +I^n)}_\bj\zbz
\end{eqnarray}
where on the multi-indices $I^\a = (i^\a_1,\dots,i^\a_n)$ the summand
$I = {\displaystyle \sum_{\a=1}^n I^\a}$ is the linear addition on
the monoid of positive integers ${\mathbb N}^n$ while
$\lam^\b_{I^\b}\zbz = {\displaystyle \prod_{r=1}^n
\Big(\prt_r\cZ^\b\zbz\Big)^{i^\b_r} }$. Moreover, in the above
expansion $\mu^r_\bj$ must be
identified with $\mu^r_\bj\,{}^{(0)}$ --see \rf{link} and \rf{2d-beltra}.

Furthermore, the symplectic structure of the space ought to provide by
virtue of \rf{mudeform} a recursive
construction for the coefficients $\mu_{\bj}^i\,{}^{(A)}$ defined in
\rf{mudeform} for the complex structure in terms of those of
Bilal-Fock-Kogan defined in \rf{muI}.  
This certainly allows to write

\begin{eqnarray}
\mu^i_\bj\zY = \sum_{|I|\geq 1} \cF^i_{(I)}\zY\, \mu^{(I)}_\bj\zbz
\end{eqnarray}
where the very complicate coefficients $\cF^i_{(I)}\zY $ depending
on the $\cG$ carry a well defined geometrical meaning.

\section{Classical $\cW_\infty$-algebra in $n$-complex dimensions}

Due to the holomorphically split expansion \rf{holphi}, the action  of
the BRS operator
$\cS$ on the theory can be parametrized by means of new ghost fields
directly obtained from this expansion. These will be intimately related to the
$\cW_\infty$-algebra. Indeed, since $\cS\cY_\a=0$, by using \rf{jetZ},
for $|A|\geq 1$, one gets the same combinatorial expansion as \rf{cG}
\begin{eqnarray}
\cS\cZ^{(A)}\zbz &=& \frac{1}{A!}\, \frac{\prt^{|A|}}{\prt\cY_A}\L\zY
\Big|_{\cY=0} = \sum_{1\leq |I|\leq|A|}
\cG^{(A)}_{(I)}\zbz \, \cC^{(I)}\zbz
\lbl{ScG}
\end{eqnarray}
where we have introduced the independent ghost fields
\begin{eqnarray}
\cC^{(I)}\zbz := \frac{1}{I!}\, \cD^{(I)}\zY \L\zY
\Big|_{\cY=0} := \prod_{k=1}^n \left( \frac{1}{i_k!}\,
\Big(\cD^k\zY\Big)^{i_k} \right) \L\zY\Big|_{\cY=0} .
\lbl{cC}
\end{eqnarray}
Note that from the very definitions, the dependence on the generalized
Bilal, Fock and Kogan pararameters can be isolated and turns out to be
coupled to the ghost $\bc{\,}^{\bi}:=\bc^{(0,\bi)}$,
\begin{eqnarray}
\cC^{(I)}\zbz = \mu_{\bi}^{(I)}\zbz \bc^{(0,\bi)}\zbz + \cdots\ ,
\qquad \qquad \bc^{(0,\bi)}\zbz :=
\frac{\prt\Lambda\zY}{\prt \obby_{\bi}}\Big|_{\cY=0},
\lbl{cCbc}
\end{eqnarray}
the full detailed expression will be given down below --see \rf{fullcCbc}.

Notably, after a tedious combinatorial calculation based upon the
commutators \rf{sd}, the BRS variations
of the ghosts defined by \rf{cC} turn out to be local (in the sense
that do not depend on the $\lambda$-fields), namely, for
$|I|\geq 1$,
\begin{eqnarray}
\hskip -1cm
\cS\cC^{(I)}\zbz &=& \sum_{k=1}^n (1-\delta_{0i_k})
\sum_{J^{(k)}\leq I^{(k)}-1_k}
\frac{(I-J^{(k)}-1_k+1_r)!}{(I-J^{(k)}-1_k)!} \times \nn\\
&&{} \lbl{sCI} \\
&&\hskip 2cm
\cC^{(I-J^{(k)}-1_k+1_r)}\zbz \prt_r \cC^{(J^{(k)}+1_k)}\zbz\nn
\end{eqnarray}
where the notation $J^{(k)}$ means $J^{(k)}=(j_1,\dots,j_k,0,\dots,0)$
(and similarly for $I^{(k)}$), $J\leq I$ is a shorthand for $j_k\leq
i_k$, $k=1,...,n$ and
\begin{eqnarray}
\frac{(I-J^{(k)}-1_k+1_r)!}{(I-J^{(k)}-1_k)!} = \left\{
\begin{array}{ll}
i_r - j_r + 1 & \mbox{ if } 1\leq r \leq k-1\\
i_k - j_k & \mbox{ if } r=k\\
i_r + 1 & \mbox{ if } k+1\leq r\leq n
\end{array} \right.
\end{eqnarray}
This formula represents the extended version to $n$ complex
dimensions of the {\em chiral} $\cW_\infty$-algebra. Indeed, let us
consider $n=1$ a complex curve which represents a bidimensional
theory built on a Riemann surface. In that case, the multi-index $I$
reduces to a simple index and for $|I|=i_1=m,\
J^{(1)}=(j_1)=j,\ r=k=1$ , the formula \rf{sCI} reduces (with
$\ell=m-j$) to that found in \cite{symplecto,complexstruct}
\begin{eqnarray}
\cS \cC^{(m)}\zbz = \sum_{\ell=1}^m \ell\, \cC^{(\ell)}\zbz \prt
\cC^{(m-\ell+1)}\zbz,
\lbl{n=1}
\end{eqnarray}
and recalled in the introduction -see \rf{n1}.
Going back to the general case, at first order $|I|=1$, we refind the usual BRS
transformations for the chiral ghosts
$\cC^i$ under diffeomorphisms of $M$~\cite{BL98}
\begin{eqnarray}
\cS \cC^i\zbz = \cC^\ell\zbz\prt_\ell\cC^i\zbz,
\end{eqnarray}
showing that diffeomorphisms are actually captured by the
$\cW_\infty$-symmetry. In order to exemplify once more \rf{sCI}, at the
second order $|I|=2$, for $1\leq i\leq j\leq n$, the multi-index
$I=(0,\cdots,0,1,0,\cdots,0,1,0,\cdots,0)$, where $1$ is at the both $i$-th
and $j$-th places will be shorthandly written as $I=(ij)$ in order to
recover a tensorial notation. With this notation, one gets
\begin{eqnarray*}
\cS \cC^{(ij)}\zbz &=& \cC^r\zbz\prt_r\cC^{(ij)}\zbz + 2
\cC^{(ii)}\zbz\prt_i\cC^j\zbz
+ 2 \cC^{(jj)}\zbz\prt_j\cC^i\zbz\\
&& +
\sum_{\scriptsize \begin{array}{c}\\[-5mm] r=1\\[-1mm] r\neq i
\end{array} }^n
\cC^{(ir)}\zbz\prt_r\cC^j\zbz\
+ \sum_{\scriptsize \begin{array}{c}\\[-5mm] r=1\\[-1mm] r\neq j
\end{array} }^n
\cC^{(jr)}\zbz\prt_r\cC^i\zbz,
\end{eqnarray*}
(no summation on $i$ and $j$ and recall that the independent ghosts
are $\cC^{(ij)}$ ). In particular, the case $i=j$ is obtained by
dividing both sides of the above equation by the symmetry factor $2$,
\begin{eqnarray*}
\cS \cC^{(ii)}\zbz &=& \underbrace{\cC^{r}\zbz\prt_r\cC^{(ii)}\zbz + 2
\cC^{(ii)}\zbz\prt_i\cC^i\zbz}_{\mbox{as in \rf{n=1} with $m=2$}}
+ \sum_{r\neq i}
\cC^{(ri)}\zbz\prt_r\cC^i\zbz
\end{eqnarray*}
where $I=(ii)$ means $2$ at the $i$-th place, a shorthand notation
saying that the multi-index entries are $i_k = 2\delta_{ki}$.
Of course, there is the complex conjugate expression to \rf{sCI} as well.

Following the BRS method recalled in the introduction, the algebra of the
$\cW_\infty$-generators in the $n$ complex dimensional case
will be obtained by duality through the corresponding BRS
functional operator \rf{BRSfunct} from 
the  BRS transformations \rf{sCI} of the chiral ghost fields themselves.
By performing this construction for the chiral ghosts, one should
directly get
a generalization to $n$ dimensions of the chiral $\cW_\infty$-algebra
\rf{Walg}. 

Accordingly, the BRS variations of the generalized Bilal-Fock-Kogan parameters
\rf{muI} can be directly computed from \rf{sCI} by using a
trick related to diffeomorphisms \cite{Ba88}, namely,
${\displaystyle \Big\{\cS,\frac{\prt}{\prt\bc^\bi}\Big\} =
\bprt_\bi,}$, together with \rf{cCbc},
\begin{eqnarray}
\cS\mu_\bj^{(I)}\zbz &=& \bprt_\bj \cC^{(I)}\zbz +
\sum_{k=1}^n (1-\delta_{0i_k})
\sum_{J^{(k)} \leq I^{(k)}-1_k}
\frac{(I-J^{(k)}-1_k+1_r)!}{(I-J^{(k)}-1_k)!} \times \nn\\
&&{} \lbl{smuI} \\
&& \hskip -2.5cm
\Big( \cC^{(I-J^{(k)}-1_k+1_r)}\zbz \prt_r \mu_\bj^{(J^{(k)}+1_k)}\zbz
- \mu_\bj^{(I-J^{(k)}-1_k+1_r)}\zbz \prt_r \cC^{(J^{(k)}+1_k)}\zbz \Big)\nn
\end{eqnarray}
By using once more the previous trick on \rf{smuI} one ends up with
the counterpart of the integrability condition \rf{integrability} in
terms of the external fields \rf{muI} for $|I|\geq 1$, and with the
aforementioned notation
\begin{eqnarray}
&& \bprt_\bi \mu_\bj^{(I)}\zbz - \bprt_\bj \mu_\bi^{(I)}\zbz =
\sum_{k=1}^n (1-\delta_{0i_k})
\sum_{J^{(k)}\leq I^{(k)}-1_k}
\frac{(I-J^{(k)}-1_k+1_r)!}{(I-J^{(k)}-1_k)!} \times \cr
&&\lbl{integrability2}\\[-1mm]
&&
\Big( \mu_\bi^{(I-J^{(k)}-1_k+1_r)}\zbz \prt_r \mu_\bj^{(J^{(k)}+1_k)}\zbz
- \mu_\bj^{(I-J^{(k)}-1_k+1_r)}\zbz \prt_r \mu_\bi^{(J^{(k)}+1_k)}\zbz
\Big) \nn
\end{eqnarray}

The chiral ghost fields $\cC^{(I)}$  admit a local
decomposition in terms of the fields \rf{muI}
which generalizes the well known conformal one
\cite{Becchi} in two dimensions. The latter has already been extended
in \cite{symplecto} for Riemann surfaces.
By definition the promissed detailed expression for \rf{cCbc} writes
\begin{eqnarray}
 \cC^{(I)}\zbz &=& \sum_{|P|,|\Qb|=0}^{|I|} \Qb!
\sum_{\scriptsize \begin{array}{c} \\[-5mm]
P+a_1J_1+\cdots + a_{|I|} J_{|I|} = I \\[1mm]
\ a_1+\cdots + a_{|I|} =
|\Qb| \end{array} }\nn\\
&&\left( \sum_{\scriptsize \begin{array}{c}
\Sb_1 + \cdots + \Sb_{|I|} = \Qb \\ |\Sb_k| = a_k,\ k=1,...,|I|\end{array} }
\prod_{k=1}^{|I|} \biggl(\frac{1}{\Sb_k!}
\Big(\mu^{(J_k)}\zbz\Big)_{\Sb_k}\biggr) \right) c^{(P,\Qb)}\zbz,
\lbl{fullcCbc}
\end{eqnarray}
where in the second summand $a_k$ enumerates (the number of
multi-indices identical to $J_k$, $J_k\neq
J_\ell$ for $k\neq \ell$ and $k,\ell = 1,...,|I|$, ($0\leq a_k \leq
|I|$), the sum goes with no repetition, and, for a given multi-index
$\Sb_k=(\ovl{s}^{(k)}_1,\dots,\ovl{s}^{(k)}_n)$
\begin{eqnarray}
\Big(\mu^{(J_k)}\zbz\Big)_{\Sb_k} :=
\Big(\mu^{(J_k)}_{\ovl{1}}\zbz\Big)^{\ovl{s}^{(k)}_1} \cdots
\Big(\mu^{(J_k)}_{\ovl{n}}\zbz\Big)^{\ovl{s}^{(k)}_n}
\end{eqnarray}
and where new independent ghost fields have been introduced by
\begin{eqnarray}
c^{(P,\Qb)}\zbz := \frac{1}{P!\,\Qb!}\, \frac{\prt^{|P|}}{\prt \bby_P}\,
\frac{\prt^{|\Qb|}}{\prt \obby_{\Qb}} \Lambda\zY\Big|_{Y=0}.
\lbl{cRS}
\end{eqnarray}
Remark that expression \rf{fullcCbc} which expresses a change of
generators for the $\cW$-symmetry is local. For instance, the case
$|I|=1$ gives
\begin{eqnarray*}
\cC^i\zbz &=& c^{(i,0)}\zbz + \mu^i_{\ovl{s}}\zbz c^{(0,\ovl{s})}\zbz
\end{eqnarray*}
which is the expression of the chiral ghost fields in terms of the
true ghost fields $c^{(i,0)}$ and $c^{(0,\ovl{s})}$ for (infinitesimal)
diffeomorphisms of $M$ \cite{BL98}, while $|I|=2$ yields
respectively
for $I=(ii)$ and $I=(ij)$, $i<j$,
\begin{eqnarray*}
\cC^{(ii)}\zbz &=& c^{(ii,0)}\zbz + \mu^i_{\ovl{s}}\zbz c^{(i,\ovl{s})}\zbz +
\Big(\mu^i_{\ovl{s}}\zbz\Big)^2 c^{(0,\ovl{s}\,\ovl{s})}\zbz+
\mu^{(ii)}_{\ovl{s}}\zbz c^{(0,\ovl{s})}\zbz\\
&& +\ \sum_{\br<\ovl{s}}
\mu^i_\br\zbz\mu^i_{\ovl{s}} \zbz c^{(0,\br\,\ovl{s})}\zbz,\\[2mm]
\cC^{(ij)}\zbz &=& c^{(ij,0)}\zbz + \mu^i_{\ovl{s}}\zbz
c^{(j,\ovl{s})}\zbz + \mu^j_{\ovl{s}}\zbz c^{(i,\ovl{s})}\zbz
+ \mu^{(ij)}_{\ovl{s}}\zbz c^{(0,\ovl{s})}\zbz\\
&& \hskip -1 cm
+\ 2 \mu^i_{\ovl{s}}\zbz \mu^j_{\ovl{s}}\zbz
c^{(0,\ovl{s}\,\ovl{s})}\zbz
+ \sum_{\br<\ovl{s}}
\Big(\mu^i_\br\zbz \mu^j_{\ovl{s}}\zbz + \mu^j_\br\zbz
\mu^i_{\ovl{s}}\zbz\Big)c^{(0,\br\,\ovl{s})}\zbz  .
\end{eqnarray*}
The ghosts $c^{(R,\Sb)}\zbz $ satisfy rather elaborate BRS transformations,
which generalize formula \rf{sCI} to the non chiral sectors. They can
be obtained either from the very definition Eq \rf{cRS} or from the
combined action of the decomposition Eq\rf{fullcCbc} and the BRS
variations \rf{sCI} and \rf{smuI}.
For $|P|+|\Qb|\geq 1$ the variations look like
\begin{eqnarray}
\cS c^{(P,\Qb)}\zbz &=& \sum_{\Sb\leq\Qb}\ \sum_{k=1}^n (1-\delta_{0p_k})
\sum_{R^{(k)}\leq P^{(k)}-1_k}
\frac{ (P-R^{(k)}-1_k+1_r)!}
{(P-R^{(k)}-1_k)!}
\times\nn\\[2mm]
&&\hspace{4cm}
 c^{(P-R^{(k)}-1_k+1_r,\Qb-\Sb)}\zbz \, \prt_r\,
c^{(R^{(k)}+1_k,\Sb)}\zbz\nn\\[2mm]
&& +\ (1-\delta_{0\Qb}) \frac{(P+1_r)!}{P!} \sum_{\Sb\leq\Qb,\
|\Sb|\geq 1} c^{(P+1_r,\Qb-\Sb)}\zbz \, \prt_r\, c^{(0,\Sb)}\zbz
\nn\\
&&\lbl{cBRS}\\[-2mm]
&& + \sum_{R\leq P}\ \sum_{\ovl{k}=1}^n (1-\delta_{0\ovl{q}_\bk})
\sum_{\Sb^{(\ovl{k})}\leq \Qb^{(\ovl{k})}-1_\bk}
\frac{(\Qb-\Sb^{(\ovl{k})}-1_{\ovl{k}}
+ 1_{\ovl{s}})!}{(\Qb-\Sb^{(\bk)}-1_\bk)!}
\times\nn\\[2mm]
&&\hspace{4cm}
c^{(P-R,\Qb-\Sb^\bk - 1_\bk +
1_{\ovl{s}})} \zbz\, \bprt_{\ovl{s}}\,
c^{(R,\Sb^\bk + 1_\bk)}\zbz \nn\\[2mm]
&& +\ (1-\delta_{0P}) \frac{(\Qb+1_{\ovl{s}})!}{\Qb!} \sum_{R\leq P,\
|R|\geq 1} c^{(P-R,\Qb+1_{\ovl{s}})}\zbz \, \bprt_{\ovl{s}}\, c^{(R,0)}\zbz,
\nn
\end{eqnarray}
and, according to the BRS technique briefly recalled in the
introduction, give rise to 
the $\cW_\infty$-structure at a non chiral level. These results
provide (in the $\zbz$ submanifold characterized by
  $\cY_\a=\bcY_{\ba}=0$ in the symplectic space) an infinite
  $\cW_\infty$-algebra  of which the first step describes
  the reparametrization invariance $\zbz\lra \ZBZz$ studied in \cite{BL98}.

Here, what is left over is the relic of the deformation process for
  $\cY_\a,\bcY_{\ba}\neq 0$ given by an infinite hierarchy of smooth changes of
local complex coordinates $\zbz\lra
(\cZ^{(A)}\zbz,\ovl{\cZ^{(A)}}\zbz)$ on the base (the $\zbz$-space) of the
symplectic space. The new $\cW_\infty$-algebra
  really encodes the behavior under symplectomorphisms of this
  hierarchy.

\sect{Towards a Lagrangian formulation}

If we wish to construct now a Lagrangian field theory whose classical limit
is invariant under this $n$-dimensional extension of a
$\cW_\infty$-algebra, it would retain the imprinting of the infinite
expansion from which the algebra is extracted, by reproducing a theory
which is badly packed in the $\zbz$ space and makes attempt to get
away in the full symplectic space.  The `classical' fields whose
dynamics serve to probe the $\cY_\a,\bcY_{\ba}\neq 0$ sector are the
generalized 
Bilal-Fock-Kogan parameters    $\mu_\bj^{(I)}\zbz$ defined in
\rf{muI}.  Indeed they are the only ``true" local fields from which the
pure gravitational theory would depend on. They are sources related to
higher spin fields as in the unidimensional complex case, see e.g. \cite{BFK}.

So from the BRS approach an infinite set of Ward
operators  $ \cW_{(I) }\zbz  $  can be obtained and from which
a classical action $\Gamma^{Cl}$ may be defined in the
vacuum sector as follows. For $|I|\geq 1$,
\begin{eqnarray}
&&\cW_{(I)}\zbz \Gamma^{Cl} = \nn\\[2mm]
&& \hskip -2cm
-\ \bprt_\bj \frac{\delta
\Gamma^{Cl}}{\mu_\bj^{(I)}\zbz}
+ \sum_{|L|\geq 1} \biggl( (i_r + \ell_r)
\Big( \prt_r \mu_\bj^{(L)}\zbz\Big) + \ell_r\, \mu_\bj^{(L)}\zbz \prt_r \biggr)
\frac{\delta
\Gamma^{Cl}}{\mu_\bj^{(I+L-1_r)}\zbz} = 0.
\end{eqnarray}
These encapsulate the first order case $|I|=|L|=1$ already treated in
\cite{BL98}.
Remark also that in order to now the $\bprt$ divergence of the higher spin
current, dual to $\mu_\bj^{(I)}$, the infinite collection of higher
spin fields must be known.

Using the usual techniques, which in two dimensional limit, lead from
 Ward identities to O.P.E. expansion \cite{BFK} we can derive a
 generalization of the ``O.P.E." algebra which would promote the
 present symplectic
 approach, since a $n$ complex dimensional short-distance product
 could generally be a difficult task to manage.

For both $|I|$ and $|J|$ greater than $0$, one obtains
\begin{eqnarray}
&&\pi \sum_{j=1}^n \int_{{\mathbb C}^{n-1}} \biggl(\! \prod_{\scriptsize
\begin{array}{c}
\ell=1\\[-1mm] \ell\neq j \end{array} }^n \!\!
\frac{d\ovl{w}^{\ovl{\ell}}\!\wedge\!dw^\ell}{2i(w^\ell-z'^\ell)} \biggr)
\frac{\delta^2 \Gamma^{Cl}}{\delta
\mu^{(I)}_\bj(w^1,\dots,z'^j,\dots,w^n,
\ovl{w}^1,\dots,\ovl{z}'^j,\dots,\ovl{w}^n)
\delta \mu^{(J)}_\bk\zbz}\Big|_{\mu=0} \nn\\
&&+\ \sum_{r=1}^n \left( \frac{i_r +j _r}{(z^r-z'^r)^2}
- \frac{i_r}{z^r-z'^r} \right)
\biggl(\! \prod_{\scriptsize \begin{array}{c}
\ell=1\\[-1mm] \ell\neq r \end{array} }^n \!\!
\frac{1}{z^\ell-z'^\ell} \biggr) \frac{\delta \Gamma^{Cl}}{\delta
\mu^{(I+J-1_r)}_\bj\zbz}\Big|_{\mu=0} =0.
\end{eqnarray}
leading to  a convolution algebra, where the directional properties
of the (short) distance limit is taken into account,  for the Green
functions generated by the generalized Bilal-Fock-Kogan fields.  Its
two dimensional limit 
gives the usual classical O.P.E. expansion.

Anyhow, due to the anomalous character of the diffeomorphism symmetry
 at the quantum level,
 we must foresee whether  this defect would be transmitted to the
 residual part of the algebra. Hence quantum corrections would be required
 to give a meaning to the theoretical model and are still under investigation.

\section{Conclusions}

It has been shown how a symplectic approach gives a strong geometrical
way of extending the notion of 
$\cW_\infty$-algebra as a symmetry arising in the one complex dimensional case
to a generic $n$  complex dimensional (compact) manifold. 

This symmetry appears from
consistent deformations of integrable complex structures in the spirit
of Kodaira-Spencer deformation. The decomposition in
terms of local quantities as the Bilal-Fock-Kogan coefficients considered
as generalized sources for higher spin 
fields  naturally emerges from the construction. However, in this
symplectic framework, a truncation process analogous to the one for
Riemann surfaces \cite{w3} from  $\cW_\infty$-algebra to a finite
$\cW$-algebra is still lacking. In particular, the latter could be of
some interest in both string and brane theories (see e.g. \cite{Castro})
where higher spin fields appear in four
real dimensions. In these theories, the fields seem to be related to
some finite $\cW$-algebra. 
However theoretical models with explicit higher spin fields still
remain to be constructed.

More generally, even if the topic ought to seem, according to the
physical context, rather technical and strongly grounded on
mathematics, we would emphasize that the important problem of a
metric or a complex structure for a physical theory embedded in a
gravitational model is far of being understood. So any little step in
that direction could give profit to discover the real of Nature  and the
intricacies of geometrical implications within the formulation of
Physical Theories.

\end{document}